# Joint Positioning and Tracking via NR Sidelink in 5G-Empowered Industrial IoT: Releasing the Potential of V2X Technology

Yi Lu[†], Mike Koivisto[†], Jukka Talvitie[†], Elizaveta Rastorgueva-Foi[†], Toni Levanen[*], Elena Simona Lohan[†] and Mikko Valkama[†]

[†] *Department of Electrical Engineering, Tampere University, Finland*

[*] *Nokia Mobile Networks, Finland*

Email : yi.lu@tuni.fi

*The fifth generation (5G) mobile networks with enhanced connectivity and positioning capabilities play an increasingly important role in the development of automated vehicle-to-everything (V2X) and other advanced industrial Internet of things (IoT) systems. In this article, we address the prospects of 5G New Radio (NR) sidelink based V2X networks and their applicability for increasing the situational awareness, in terms of continuous tracking of moving connected machines and vehicles, in industrial systems. For increased system flexibility and fast deployments, we assume that the locations of the so-called anchor nodes are unknown, and describe an extended Kalman filter-based joint positioning and tracking framework in which the locations of both the anchor nodes and the target nodes can be estimated simultaneously. We assess and demonstrate the achievable 3D positioning and tracking performance in the context of a realistic industrial warehouse facility, through extensive ray-tracing based evaluations at the 26 GHz NR band. Our findings show that when both angle-based and time-based measurements are utilized, reaching sub-1 meter accuracy is realistic and that the system is also relatively robust against different node geometries. Finally, several research challenges towards achieving robust, high-performance and cost-efficient positioning solutions are outlined and discussed, identifying various potential directions for future work.*

## Introduction and Motivation

Positioning, navigation, and location-aided communications techniques are central technical enablers in the emerging autonomous systems with connected vehicles and industrial Internet-of-Things (IoT) [1]. On top of the advanced communications features, the fifth generation (5G) networks can facilitate high-accuracy positioning service owing to the availability of the wide bandwidths and the beamforming capabilities [2]. However, deploying full-coverage 5G systems in versatile industrial environments may not be feasible, particularly at the millimeter wave (mmWave) bands with largely reduced cell sizes. As part of the 5G New Radio (NR) specifications, the so-called sidelink technology [3] allows to establish ad-hoc networks among the NR user equipment (UE) and the automated vehicles. This allows for improved network coverage and enhanced vehicle-to-everything (V2X) connectivity in complex industrial environments while can also provide new opportunities for efficient localization and tracking – a topic that is addressed, discussed and demonstrated in this article.

In the considered NR sidelink based V2X scenario, illustrated in Fig. 1, we differentiate between the so-called anchor nodes and the target nodes depending on the availability of the access link towards the NR base station. Furthermore, for improved system flexibility and in order to cope with the challenging conditions of different industrial premises [4], we assume that the anchor nodes can be deployed dynamically, on an on-demand basis. This allows for flexible relocation and re-deployment of the anchors, such as aerial vehicles or high-end terminals as local hotspots, in order to optimize, e.g., the radio coverage and line-of-sight (LoS) conditions between the anchors and the target nodes in dynamic industrial environments. However, this also implies that the locations of such dynamically deployed anchors are not precisely known, which in turn complicates the positioning and tracking of the actual industrial target nodes. To address this challenge, an advanced joint positioning and tracking framework can be devised where the locations of both the anchor nodes and the target nodes are simultaneously estimated and tracked. Such approach allows for creating increased location awareness in advanced industrial IoT systems, and thereon facilitates, e.g., efficient beam-forming, handover and radio resource management functionalities. Additionally, the high-accuracy tracking of the industrial vehicles and machines can significantly improve the overall situational awareness, thus enhancing the efficiency and safety of the industrial facility [4].

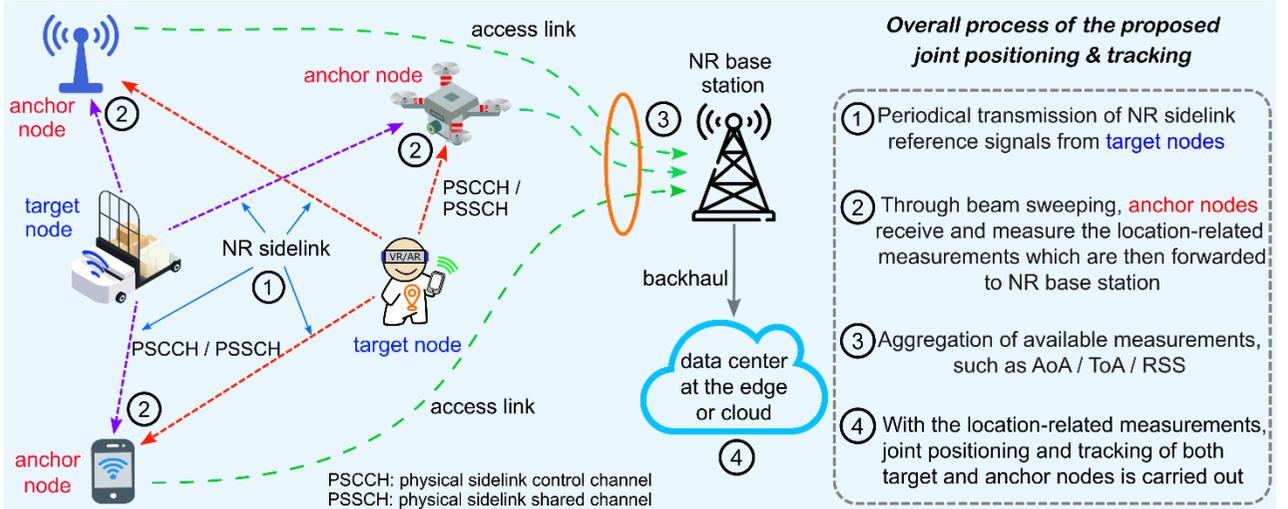

**Figure 1**: A system-level overview of the NR sidelink based ad-hoc V2X network with dynamically deployed anchor nodes and actual industrial target nodes. The joint positioning and tracking of anchors and targets builds on the sidelink-based measurements, aggregated and processed on the network side.

In this article, we describe, demonstrate and assess the performance of an advanced joint positioning and tracking framework in the above-described NR sidelink based industrial V2X network context, operating at the 26 GHz mmWave band. The approach builds on a network-centric joint positioning and tracking engine, aggregating and processing the measurements obtained by the anchor nodes. Both delay-based and angle-based measurements are considered, building on the standardized 5G NR sidelink reference signal (SL-RS) transmissions between the target and the anchor nodes. For accurate modelling and performance assessments, extensive ray-tracing based evaluations are provided comprising an accurate digital map of a realistic industrial warehouse. Furthermore, the achievable positioning and tracking accuracy under different numbers of the involved anchor and target nodes as well as the impact of the node geometry and different involved measurements are studied and demonstrated. Finally, we identify and discuss the potential open research challenges in achieving high-accuracy, cost-efficient and robust positioning solutions in 5G-empowered industrial systems.

## NR Sidelink based ad-hoc Networks for Joint Positioning and Tracking

### System Descriptions and Fundamentals

Following the 3rd generation partnership project (3GPP) standardization regarding the so-called proximity-based services [5], an industrial IoT system capitalizing the 5G NR sidelink technology at mmWave frequencies is considered. The visualized V2X network that is illustrated on the left-hand side of Fig. 1 consists of a set of IoT nodes (i.e., vehicles and devices), NR base station and an edge/cloud server acting as the management and control entity. Among all the active IoT nodes, the anchor nodes have direct radio connection with the NR base station via access link, while communicating with the target nodes through the NR sidelink. The target nodes are the industrial devices, such as robots, industrial vehicles or augmented/virtual reality glasses, while the anchor nodes can be, e.g., high-end smart phones or aerial vehicles allowing them to be deployed in a dynamic manner, flexibly providing extended coverage and enhanced capacity in the industrial facility. Furthermore, there is no particular limitations for the exchanging role of the two types of involved IoT nodes. That is, the anchor (target) nodes can act as the target (anchor) nodes depending on the situation and system configurations.

The main procedures of the sidelink-based joint positioning and tracking framework, illustrated also in Fig. 1, can be described and summarized as follows:

1. The target nodes that act as the transmitting UEs periodically transmit the sidelink channel state information reference signals (SL CSI-RS) via scheduled physical layer channels, such as physical sidelink control channel (PSCCH) and physical sidelink shared channel (PSSCH) [6];
2. At the receiving side, the anchor nodes collect the location-related measurements through a beam sweeping process [7] which will be further discussed in Section II-B. Thereafter, the obtained measurements are forwarded to NR base station for further processing;
3. At the base station side, all the available measurements are aggregated accordingly. Data association can also be involved to relate each LoS / non-line-of-sight (NLoS) measurement with the corresponding target-anchor pair;
4. With the associated location-related measurements, joint positioning and tracking of both targets and anchors is carried out at the network side. Herein, we apply an extended Kalman filter (EKF)-based solution which will be discussed in technical terms in Section II-C.

From the system feasibility perspective, the considered network-centric approach efficiently exploits the computational power offered by edge computing. Additionally, the approach allows the estimated location information to be directly utilized at the network side for the optimization of dynamic spectrum sharing, radio resource allocation, mobility management and scheduling functionalities, for instance.

With regards to the sidelink operations and resource allocation, a direct-device-discovery mechanism as specified in the context of NR V2X [8] is assumed to be implemented among all the nodes within the sidelink coverage range to ensure a pervasive wireless connectivity. In addition, the physical layer resource allocation can be scheduled either by the NR base station or by the transmitting target nodes in the contention-based fashion. The specified physical channels, such as PSCCH and PSSCH are typically scheduled in a time-division-multiplexed manner to optimize the energy efficiency and (end-to-end) latency [9].

In general, if the sidelink connections among all the IoT nodes – including not only the target-to-anchor, but also the target-to-target and the anchor-to-anchor connections – were considered, the achievable positioning accuracy can be high, however, at the expense of high complexity, interference and latency. Thus, in order to relax the overall complexity and potential interference within the network, only the sidelink between the anchor nodes and the target nodes are established and utilized in this work, as illustrated by the purple and red dashed arrows in Fig. 1. To this end, also the joint positioning and tracking entity, described in Section II-C, builds only on the measurements between target nodes and anchor nodes.

### Location-related Measurements

In general, the essence of radio positioning is to solve a collection of non-linear equations given the available location-related measurements like, the time of arrival (ToA) and the angle of arrival (AoA). The accuracy of the employed measurements, therefore, directly affects the final positioning performance. From the system perspective, the total number of available measurements depends on the number of involved target nodes as well as the deployed anchor nodes, whereas the attainable measurement accuracy varies according to the signal characteristics and the corresponding specific estimation methods.

In the considered NR sidelink based ad-hoc V2X network, beamformed measurements are one key technical ingredient. To this end, we consider a beam sweeping approach for ToA and AoA measurements stemming from the defined and regulated procedures for NR beam management [7]. As the first two steps of the overall scheme illustrated in Fig. 1, the industrial target nodes periodically transmit the standardized NR SL CSI-RS [10], which the anchors are then receiving and measuring in a beam-based manner. The AoA measurements are determined by observing the power differences between the separate receive beams in the involved beam sweeping process, also known as the beam reference signal received power (B-RSRP). The ToAs, in turn, are estimated by calculating the phase differences between the different sub-carriers of the received orthogonal frequency division multiplexing (OFDM) signal.

However, there are certain challenges related to reliable acquisition and utilization of ToA and AoA measurements. In terms of ToA, there typically exists a time-varying clock offset between any two agents. Without proper synchronization, ToA measurements do not provide good positioning accuracy. Such a problem can in general be overcome by performing time difference of arrival (TDoA)-based positioning with synchronized anchor nodes. Alternatively, a joint positioning and synchronization scheme can be applied to provide reliable positioning as well as to simultaneously estimate and track the involved synchronization errors [11]. As for AoA, one common problem stems from the uncertainty of the orientations in the involved antenna arrays. Unlike the conventional base stations whose locations and array orientations can be typically assumed to be known at the network side, such information is not directly available with the envisioned anchor nodes, such as on-demand aerial vehicles. With proper formulation of the Bayesian estimation filters, both the target node locations and the anchor node orientation errors can be jointly estimated [12]. Alternatively, an inertial measurement unit (IMU) can be employed to provide an accurate orientation estimate.

### Joint Positioning and Tracking

With the location-related measurements acquired via sidelink interface, the actual joint positioning and tracking framework builds on an EKF, motivated by its flexibility to deal with both linear and non-linear measurement models and by its relatively low computational complexity compared to other Bayesian filters, such as particle filters [13]. Specifically, the joint positioning and tracking can be achieved using an EKF by recursively finding the *posterior distribution* of the states of target nodes and anchor nodes based on the available measurements and prior knowledge on their locations

$$P(s[n]|y[1:n]) \propto P(y[n]|s[n]) \cdot P(s[n]|y[1:n-1]) \quad (1)$$

where $\mathbf{s} = \left[\mathbf{s}_T^T, \mathbf{s}_A^T\right]^T$ refers to the joint state vector that consists of the targets' states and anchors' states, which are denoted as $\mathbf{s}_T$ and $\mathbf{s}_A$. As a Bayesian framework, the joint positioning and tracking functionality can be realized by extracting the location information at the current time instant $n$ from the estimated AoA and ToA measurements as well as the prior knowledge that is acquired from the previous time instants [13]. In this work, we particularly focus on the moving target nodes and static anchor nodes, thus, the locations, velocities and accelerations of the target nodes are included in their state vector, whereas for the anchor nodes, only the locations are considered. For detailed descriptions of the EKF formulations and measurement models, the readers can refer to [14]. Furthermore, positioning of dynamic anchor nodes can also be achieved by adding the corresponding velocities and accelerations in the state vector and the corresponding state transition model of the applied EKF.

It is also noteworthy that, as a recursive estimator, the EKF normally takes a few iterations (i.e., time instants) to eventually converge towards the true locations of both target and anchor nodes. Such convergence time depends on the accuracy of utilized location-related measurements as well as the initial uncertainty of all the involved nodes in the state vector. The convergence behavior of the considered joint positioning and tracking approach can be observed in a complementary multimedia demonstration available at https://research.tuni.fi/wireless/research/positioning/NR_sidelink_positioning/. In addition, the computational efforts grow with the number of involved nodes, both targets and anchors. Nevertheless, the involved processing is implementation feasible for modern edge computing systems.

## Performance Evaluations and Analysis

### System Scenario and Initialization

The performance evaluation of the joint positioning and tracking framework is conducted in an indoor industrial warehouse facility with an overall open space of 70 m×25 m×18 m, as depicted in Fig. 2. For realistic propagation modeling of the sidelink transmissions and their beam-based reception, full ray tracing calculations are carried out with Wireless InSite, incorporating an accurate digital map of the industrial facility and the different objects therein. Specifically, the anchor nodes are deployed in two representative geometric relationships, which we refer to as the collinear set and non-collinear set. Such configuration is designed to investigate the impact to the positioning performance under different geometric conditions that are generally defined by the geometric dilution of precision (GDoP). As illustrated in Fig. 2, the collinear setup (illustrated in red) implies a challenging GDoP, whereas the non-collinear setup (illustrated in green) yields a beneficial GDoP. Such impact of the geometry on the positioning and tracking performance is assessed and demonstrated in the numerical results.

Different industrial target nodes move along specific random waypoint based 3D trajectories through the open space with velocities of 1.0-1.2 m/s. Furthermore, the overall length of target nodes' trajectories is set to 5 min, resulting in 5000-sample trajectories for a 100 ms EKF update interval. The location-related measurements at the anchor nodes build on the SL CSI-RS transmissions [10] of the target nodes with 100 ms periodicity (i.e., the same period as the EKF update interval), which are omni-directionally transmitted with 0 dBm power over 100 MHz bandwidth at the 26 GHz carrier frequency, while the NR network subcarrier spacing is assumed to be 60 kHz with 1620 active subcarriers in total. When multiple target nodes exist, the reference signal transmissions can be multiplexed in time or frequency, or through orthogonal codes, such that anchors can measure them without mutual interference. Finally, the receiver sensitivity of the anchor nodes is assumed to be −90 dBm, thus, whenever the received signal power is below the sensitivity threshold, the sidelink connection cannot be established.

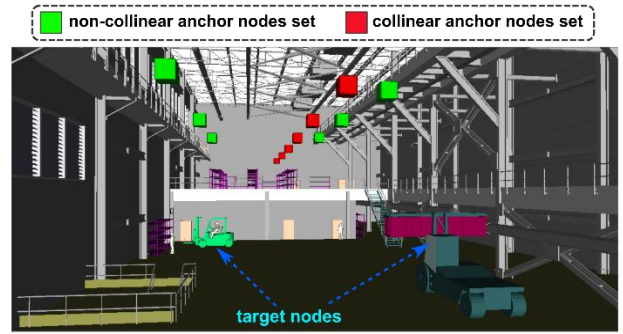

**Figure 2**: *A ray-tracing featured indoor industrial warehouse, from Wireless InSite, containing several target nodes and anchor nodes. Full ray-tracing data available at: https://dx.doi.org/10.21227/cms9-de61*

Both the periodicity of the measurements and the update interval of EKF are selected to be 100 ms, allowing for realistic beam-based sidelink measurement patterns as well as time to communicate and process the measurements. It is important to note that a shorter periodicity/update interval leads to more frequent beam sweeping and high computational efforts of the anchor nodes, whereas a longer interval may in general cause unacceptable latency for certain V2X use cases, such as autonomous driving and platooning. As discussed in Section II-C, the EKF state vector contains the 3D locations of the anchor nodes and the 3D locations, velocities and accelerations of the mobile target nodes. The system initialization builds on anchor location estimates, that can in practice be provided by the base station through uplink SRS measurements. Furthermore, the initial locations of the targets can be estimated through the delay or angle measurements via, e.g., ordinary batch least-squares at the first time instant. Moreover, despite the LoS paths between each pair of target and anchor nodes typically exists in the considered industrial scenario, it can also happen that the LoS paths are blocked by different objects in the environment. In such cases, the location-related measurements of the NLoS anchor node are omitted in the estimation process if the target node location while the output of the EKF's prediction step is considered as the location estimates for the anchor nodes.

### Numerical Results and Analysis

First, we investigate the accuracy of the location-related measurements, namely ToA/TDoA (converted to meters) and AoA (converted to degrees), serving as the inputs of the joint positioning and tracking framework. As can be observed from Fig. 3, the delay measurements (ToA and TDoA) possess a very high accuracy, that is, roughly 95% of the errors are smaller than 0.2 m. This is stemming from the 100 MHz bandwidth that provides essentially sufficient resolution to distinguish the LoS path from other multipaths. In the angular domain, the estimation errors of both elevation and azimuth AoA undergo a similar level of accuracy with 95% of the errors being less than 2 degrees. This is due to the fact that the beamwidths, i.e., the spatial angular resolution, in the elevation and azimuth planes are nearly the same since an 8×8 uniform rectangular array (URA) antenna panel is considered for all the anchor nodes. We note that the AoA accuracy can be further improved using larger antenna array with a finer angular resolution for beam sweeping and AoA estimation.

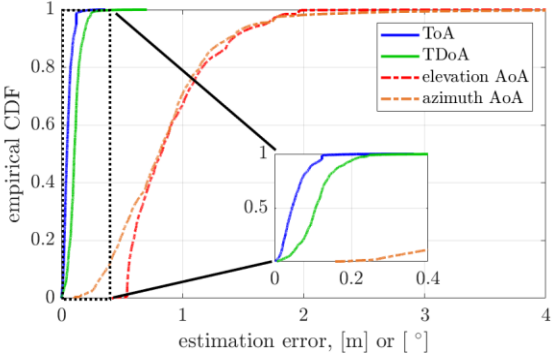

**Figure 3**: *The empirical cumulative density functions of the estimation errors of both the delay-based (ToA/TDoA) and the angle-based (AoA) location-related measurements corresponding to the beam-based reception of the sidelink signals.*

Next, we assess and evaluate the actual joint positioning and tracking performance, considering different combinations of the available delay and angle measurements under the two representative geometries, i.e., collinear and non-collinear deployments illustrated in Fig. 2. It is noted that in all the forthcoming results, the positioning performance is measured and quantified jointly for all the involved nodes – the anchors and the targets. It is further noted that the trends that can be observed from such jointly quantified results are essentially identical to those that can be obtained when separately plotting the tracking performance of either the target nodes or the anchor nodes.

The achieved performance results as well as the corresponding performance thresholds of 1 m for 2D and 0.2 m for vertical plane, noted in [15], are collected and shown in Fig. 4. The overall accuracy is characterized by the median value marked by the dark line around the center of each performance bar. Moreover, the sub-meter and sub-0.2 meter thresholds are plotted therein as a measure of the reliability of the obtained location awareness.

The 2D performance with 6 anchor nodes and with different combinations of the available measurements are given in the left part of Fig. 4. We see that both the median value of the positioning error and the probability of sub-1 meter accuracy when utilizing delay measurements (ToA or TDoA) suffer largely from the collinear deployment (red and dark red boxes) because of the corresponding challenging GDoP. However, the performance of the system applying delay measurements improves substantially when the GDoP becomes more favorable through the non-collinear deployment (green and dark green boxes). In such case, the two-target ToA-based system achieves the median 2D accuracy of 0.5 m, which is already rather close to the positioning performance utilizing ToA+AoA and TDoA+AoA.

The corresponding vertical domain performance is illustrated on the right part of Fig. 4. It can be observed that, even under non-collinear scenario, the delay-based systems perform rather poor, with the median positioning error ranging from 2 m to 8 m – i.e., substantially larger than the targeted 0.2 m threshold stated in [15]. On the contrary, the AoA-based system achieves an accuracy level close to the target threshold, showing better performance in vertical direction compared to that in 2D. Furthermore, when utilizing both the delay and angle measurements (i.e., ToA+AoA and TDoA+AoA), the vertical performance remains at the similar excellent level as the corresponding 2D performance. In general, with the combined measurements, the sub-0.2 meter vertical accuracy and the sub-1 meter 2D accuracy are achieved in more than 85% and 90% of the cases, respectively. Examining further the results in Fig. 4, it is important to note that the performance in both 2D and vertical planes are in general better when there exists two target nodes (the light color bars), compared to the single target node case (the dark color bars). Moreover, the performance gain by adding one more target node is significant when building the positioning and tracking only on a single measurement (ToA, TDoA or AoA), while the gain is less pronounced when multiple measurements are fused together.

Pursuing to further understand the impacts of the amount of the involved nodes, we explore the positioning performance as a function of the numbers of the target nodes and anchor nodes. Aiming at achieving the best possible performance, we consider TDoA+AoA based joint positioning and tracking under the same industrial environment, shown in Fig. 2, while now assuming randomly dropped anchor locations in the considered 3D space. The corresponding performance results in terms of the probability of achieving sub-1 meter accuracy in 3D are plotted in Fig. 5. Overall, the worst performance can be observed at the bottom row when only two anchor nodes are deployed. The corresponding performance is less than 60% regardless of the total number of target nodes. Moving towards upper rows, the sub-1 meter probability increases since more anchor nodes are involved in the network. The performance gain is significant when the number of anchor nodes changes from two (40–58% probability for sub-1 meter 3D positioning accuracy) to three (69–80%), and from three to four (more than 85%). Thereafter, more anchor nodes do not anymore yield significant further improvements, the best performance being 87% with two target nodes and six anchor nodes in total. From the joint positioning and tracking point of view, we can conclude that reliable location awareness (larger than 76% 3D sub-1 meter probability) can be achieved with four, five or six anchor nodes, independent of the number of target nodes, when deploying ToA/TDoA+AoA measurements. When the network consists of more than one target node and more than three anchor nodes – a typical situation for V2X – the performance of joint positioning and tracking with high number of target nodes (e.g., eight) becomes slightly worse compared to the performance when dealing with less target nodes (e.g., two, or three). The main reason lies in the fact that more target nodes bring higher dynamics to the V2X network, yielding a challenge joint positioning and tracking situation. Lastly, it is noted that while the majority of the examples and discussions in the article consider indoor scenarios, there are no technical limitations in applying similar technological approach also in outdoor environments.

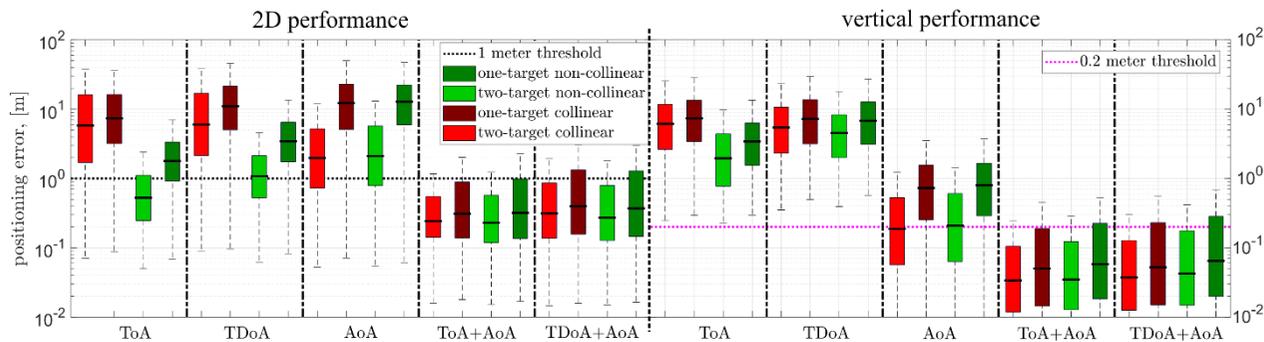

**Figure 4**: *Achieved positioning performance with the EKF-based joint positioning and tracking framework under collinear and non-collinear deployments with six anchor nodes and one or two target nodes.*

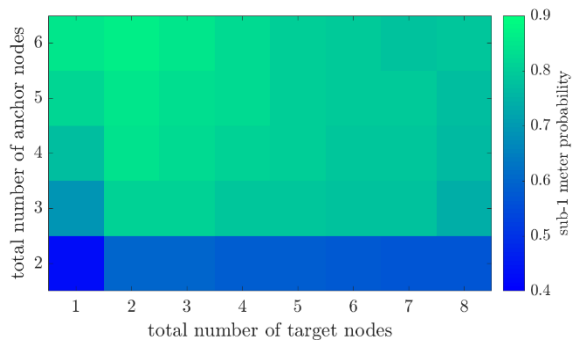

**Figure 5**: *The probability of sub-1 meter 3D positioning accuracy as a function of the numbers of target nodes and anchor nodes when utilizing TDoA and AoA measurements in the proposed joint positioning and tracking framework.*

## Challenges and Further Research Directions

We next identify and discuss selected challenges and potential future research directions to achieve high-accuracy, cost-efficient and robust positioning and tracking solutions in industrial V2X systems, illustrated also in Fig. 6.

**Deploying both fixed and dynamic anchors:** The deployment of unplanned and movable dynamic anchors discussed in the article comes with the expense of an increased system-level management complexity. Moreover, such dynamic anchors may suffer from short operational periods because of, e.g., a limited battery capacity of aerial vehicles. Additionally, the antenna systems are generally more limited in terms of, e.g., number of antenna elements, implying less accurate AoA estimation capabilities compared to actual base stations. As a solution, the joint deployment of fixed and dynamic anchors can form multi-connectivity for enhanced communication and positioning in different industrial IoT scenarios, calling for further research in terms of system design and optimization.

**Accurate vertical estimation with and without AoA:** While 2D positioning can generally be sufficient for several timely use cases, such as modern high-speed train and pedestrian navigation, the vertical accuracy is essential in various industrial verticals in which the vehicles/robots are moving along trajectories in 3D space. In this work, we have seen that the vertical positioning accuracy reaches sub-meter level when AoA measurements are utilized, either individually or fused with ToA measurements. However, to reliably measure the AoA one must solve and optimize several practical issues such as the array orientation and beam-resolution, while also be able to operate efficiently in NLoS scenarios. Therefore, dedicated research efforts are needed to ensure a high vertical accuracy in various industrial use cases, with and without AoA measurements, and possibly further complemented with IMU and other relevant onboard sensors.

**Smart management and optimization of resources, system geometry and positioning accuracy:** The positioning performance of any radio-based system depends primarily on the available radio resources, such as signal bandwidth and transmission power, with direct impact on the accuracy of the obtained location-related measurements and the corresponding positioning. However, with the relative geometrical relationships between the targets and the anchors being taken into account, smart resource management and geometry optimization schemes deserve further efforts, to balance the performance among the decisive parameters – a fundamental sweet spot.

**Positioning and tracking in LoS-obstructed environments:** With the growing complexity of indoor (e.g. multi-floor industrial warehouse) and outdoor (e.g. harbor or container port area) environments, the radio propagation between transmitting and receiving devices becomes increasingly complex, especially with the dynamic blockers. As a result, there is an increasing likelihood for the LoS path being completely obstructed, yielding challenging NLoS scenarios. Thus, novel positioning, tracking and mapping frameworks employing both the multipath components and the relay paths from the intelligent surfaces are seen to form an intriguing open research space to synthesize the location and environmental awareness, towards an intelligent positioning and mapping system.

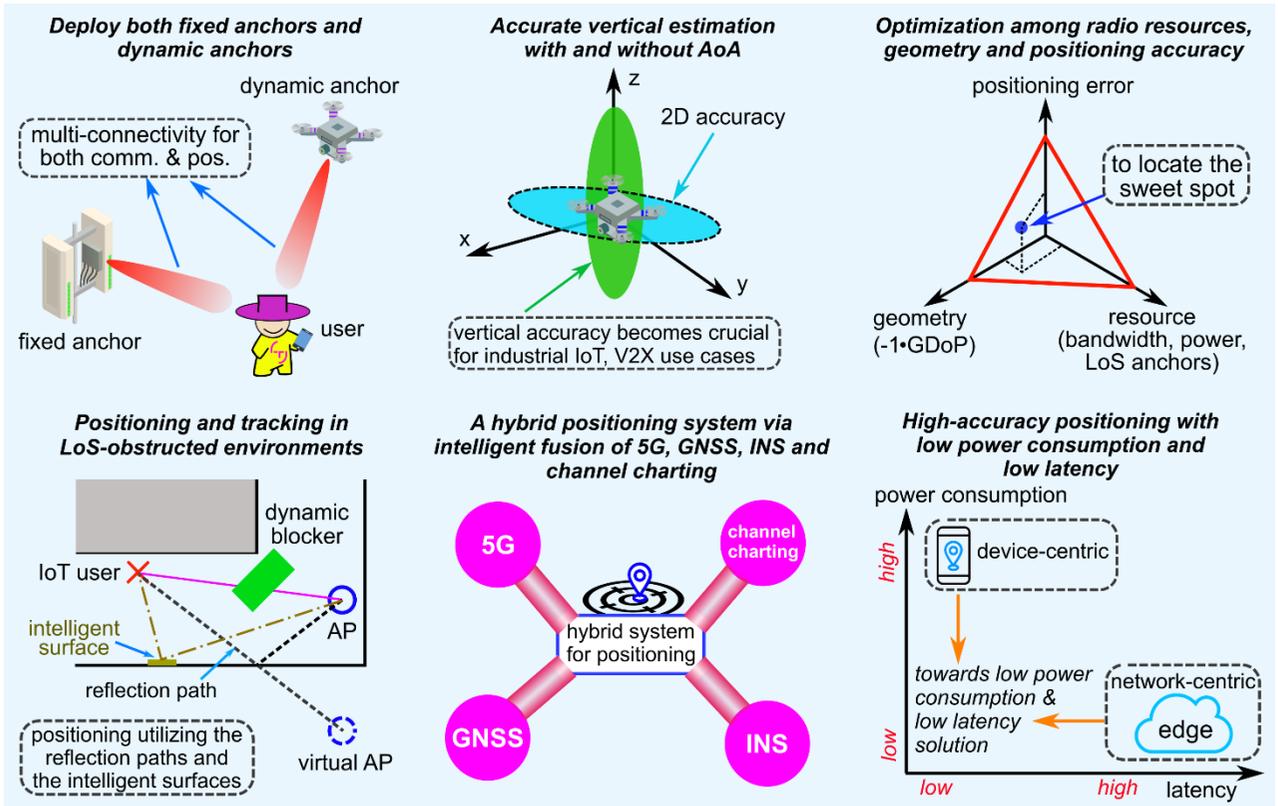

**Figure 6**: *Potential research directions towards high-accuracy, robust, and cost-efficient positioning solutions.*

**Hybrid positioning through fusion of 5G, global navigation satellite systems (GNSS), inertial navigation system (INS) and channel charting:** Today, there are several positioning system and technologies for location estimation in different environments. However, the performance when applying only one type of system can be hardly satisfactory in all scenarios. Therefore, hybrid positioning schemes based on the fusion of 5G cellular, GNSS+INS and channel charting are to be studied and developed towards a universal solution for robust positioning of aerial or ground vehicles in urban, rural and indoor scenarios – a topic that is broad and deserves substantial research efforts.

**High-accuracy positioning with low power consumption and low latency:** For a device-centric framework, high-accuracy positioning may easily infer high power consumption since both the measurements and their processing are carried out by the device. Alternatively, in the network-centric approach considered also in this article, the device power consumption is largely reduced through offloading the processing of the location-related measurements to the network edge/fog or cloud servers. Such entities commonly possess at least an order of magnitude larger computational capabilities, however, at the expense of clearly larger latencies. Hence, the investigation of the trade-offs between the power consumption and the latency for high-accuracy positioning is generally one viable research direction.

## Conclusions

Aiming at unlocking the full potential of V2X communications for 5G-empowered industrial IoT, we described, discussed and assessed a joint positioning and tracking framework utilizing the NR sidelink technology. Building on beam-based measurements acquired via NR SL CSI-RS, a network-centric extended Kalman filter based solution was envisioned to estimate and track the locations of both the anchor nodes and the actual industrial target nodes simultaneously.

The achievable positioning performance was evaluated utilizing both delay and angle-based measurements via extensive ray-tracing simulations at 26 GHz in a realistic industrial environment. The numerical results demonstrated, among others, that by incorporating the angle measurements into the positioning and tracking system, the positioning performance becomes robust to different geometry setups and that reliable location awareness can be attained in both 2D and vertical domains. Furthermore, a clear performance gain was observed when increasing the number of target nodes for low number of anchors. Overall, the results indicate that in the considered industrial warehouse type of facility, high positioning and tracking performance can be obtained even with a relatively low number of anchor nodes, reflecting a cost-efficient and flexible system.

Finally, several research challenges and potential further directions for high-accuracy, robust and cost-efficient positioning in wireless industrial IoT systems were discussed. Moreover, with proper combination of efficient positioning and sensing techniques, the obtained situational awareness of the node locations as well as the surroundings and environment can be employed to facilitate improved industrial efficiency and safety.

## Acknowledgements

This work was financially supported by the Academy of Finland, under the projects ULTRA (328226, 328214) and FUWIRI (319994), and by the Finnish Funding Agency for Innovation under the project 5G VIIMA.

**Biographies**

*Yi Lu* is a Researcher at Tampere University. His research interests include network-centric positioning system and positioning-aided communications in mmWave mobile networks.

*Mike Koivisto* is a Researcher at Tampere University. His research interests include positioning, with an emphasis on positioning and the utilization of location information in mobile networks.

*Jukka Talvitie* is a University Lecturer at Tampere University, Finland. His research interests include signal processing for wireless communications, and network-based positioning methods with particular focus on 5G networks.

*Elizaveta Rastorgueva-Foi* is a Researcher at Tampere University. Her research interests include positioning and location-aware communications in mmWave mobile networks with a focus on user mobility.

*Toni Levanen* is a SoC Specialist at Nokia Networks, Tampere, Finland. His research interests include physical layer design and SoC development for 5G NR and beyond mobile networks.

*Elena Simona Lohan* is a Full Professor at the Electrical Engineering unit, Tampere University, Finland and the coordinator of the MSCA EU A-WEAR network. Her current research interests include wireless location techniques, wearable computing, and privacy-aware positioning solutions.

*Mikko Valkama* is a Full Professor and Department Head of Electrical Engineering at Tampere University, Finland. His research interests include radio communications, radio systems and signal processing, with specific emphasis on 5G and beyond mobile networks.